\documentclass[12pt,notoc]{JHEP3}

\usepackage{amsmath,amssymb,euscript,array,cite}

\input{epsf}
\usepackage{epsfig}

\def\a{\alpha}

\def\c{\gamma}
\def\d{\delta}

\def\l{\lambda}
\def\m{\mu}
\def\n{\nu}

\def\s{\sigma}
\def\t{\tau}

\def\w{\omega}

\def\Dbarslash{\,\,{\raise.15ex\hbox{/}\mkern-12mu {\bar D}}}
\def\Dslash{\,\,{\raise.15ex\hbox{/}\mkern-12mu D}}
\def\delslash{\,\,{\raise.15ex\hbox{/}\mkern-9mu \partial}}
\def\delbarslash{\,\,{\raise.15ex\hbox{/}\mkern-9mu {\bar\partial}}}

\newcommand{\EQ}[1]{\begin{equation} #1 \end{equation}}

\newcommand{\SP}[1]{\begin{equation}\begin{split} #1
\end{split}\end{equation}}

\title{Causality and Micro-Causality in Curved Spacetime}
\author{Timothy J. Hollowood and Graham M. Shore\\
Department of Physics,\\ University of Wales Swansea,\\
Swansea, SA2 8PP, UK.\\
E-mail: {\tt t.hollowood@swansea.ac.uk}, {\tt g.m.shore@swansea.ac.uk}}

\abstract{We consider how causality and micro-causality are realised in
QED in curved spacetime. The photon propagator is found to exhibit novel 
non-analytic behaviour due to vacuum polarization, which invalidates the
Kramers-Kronig dispersion relation and calls into question the validity
of micro-causality in curved spacetime. This non-analyticity is 
ultimately related to the
generic focusing nature of congruences of geodesics in curved
spacetime, as implied by the null energy condition, and the existence of 
conjugate points. These results arise from a calculation of the
complete non-perturbative frequency dependence of the vacuum polarization
tensor in QED, using novel world-line path integral methods together with
the Penrose plane-wave limit of spacetime in the neighbourhood of a null
geodesic. The refractive index of curved spacetime is shown to exhibit
superluminal phase velocities, dispersion, absorption (due to 
$\gamma\to e^+e^-$) and bi-refringence, but we demonstrate that the
wavefront velocity (the high-frequency limit of the phase velocity)
is indeed $c$, thereby guaranteeing that causality itself is respected.}

\begin{document}

\section{Introduction}

The purpose of this letter is to consider how causality and micro-causality 
are realised in quantum field theory in curved spacetime in the light of
the discovery of novel non-analytic behaviour in the photon propagator
due to vacuum polarization in QED \cite{us}.

These questions have arisen through the resolution of a long-standing puzzle
in ``quantum gravitational optics'' \cite{Shore:2003zc,Shore:2007um},
{\it viz.\/}~how to reconcile the fact that the low-frequency phase velocity
$v_{\rm ph}(0)$ for photons propagating in curved spacetime may be 
superluminal \cite{Drummond:1979pp} with the requirement of 
causality\footnote{In fact, the question of whether causality could be 
maintained in curved spacetime even if the wavefront velocity exceeds $c$
is more subtle and involves the general relativistic notion of ``stable
causality'' \cite{HawkingEllis,Liberati:2001sd}. See ref.~\cite{Shore:2003jx}
for a careful discussion.} that the wavefront velocity $v_{\rm wf} =
v_{\rm ph}(\infty)$ should not exceed $c$. 

The reason why this is a problem is related to the Kramers-Kronig dispersion
relation \cite{Kramers,Kronig}. In terms of the
refractive index $n(\w)$, where $v_{\rm ph}(\w) = 1/ \text{Re}~n(\w)$  
(setting $c = 1$), this is
\EQ{
\text{Re}\,n(\infty)-\text{Re}\,n(0)=-\frac2\pi\int_0^\infty
\frac{d\omega}\omega\,\text{Im}\,n(\omega)\ .
\label{aa}
}
Provided $\text{Im}~n(\w) > 0$, as required by unitarity in the form of the 
optical theorem, this implies $\text{Re}\,n(\infty) < 
\text{Re}\,n(0)$ and hence $v_{\rm ph}(\infty)>
v_{\rm ph}(0)$. The fundamental assumption in the derivation of
eq.~\eqref{aa} is that $n(\w)$ is analytic in the upper-half complex
$\w$ plane, which is generally presented (see, {\it e.g.\/}~ref.~\cite{Weinberg}) in flat spacetime as a direct consequence of micro-causality. 

\begin{figure}[ht] 
\centerline{\includegraphics[width=2.5in]{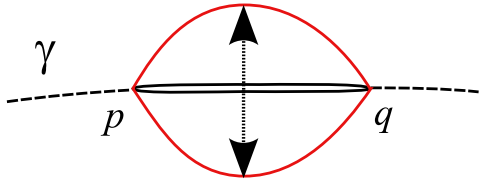}}
\caption{\footnotesize Points $p$ and $q$ on a geodesic $\gamma$ are 
{\it conjugate\/} if they can be joined---at least in an infinitesimal
sense---by a family of geodesics
close to $\gamma$. The existence of conjugate points means
that the classical saddle-point solution with the photon connected at
$p$ and $q$ to a degenerate $e^+e^-$ loop squashed
onto the geodesic $\gamma$ is deformable into a non-degenerate 
loop as shown.}\label{pic17} 
\end{figure}

The resolution of this apparent paradox is that the Kramers-Kronig dispersion
relation does {\it not\/} hold in the form \eqref{aa}. 
We find that $n(\w)$ develops branch-point singularities 
on the imaginary axis in the upper-half plane.
The consequent modification of eq.~\eqref{aa} then allows $\text{Re}\,
n(\infty) > \text{Re}\,n(0)$ 
and we find by explicit calculation of the full non-perturbative frequency
dependence of the refractive index that $n(\infty)$ is indeed equal to 1
and the wavefront velocity itself is $v_{\rm wf} = c$. Remarkably, we find
that these unusual analyticity properties can be traced very directly to the
focusing property of null congruences and the existence of conjugate
points, which are, in turn, a consequence of the null energy
condition. Conjugate points are points in spacetime that are joined
by a family of geodesics: at least in an infinitesimal sense, see
fig.~(\ref{pic17}). Such
infinitesimal deformations are associated to zero modes that lead
directly to the branch points on the imaginary axis.

The refractive index in QED is determined by the vacuum polarization tensor 
$\Pi_{\m\n}(k)$. On-shell, and in a basis diagonal with respect to
the polarizations $\hat\varepsilon_i$ ($i = 1,2$), the relation is
\EQ{
n_i(\w) = 1 +{1\over\w^2} \Pi_{ii}(\w)\ .
\label{ab}
}
We have, for the first time, evaluated the complete non-perturbative 
frequency dependence of 
the vacuum polarization $\Pi_{ij}(\w)$ in QED in curved spacetime by combining
two powerful techniques: ~(i) the world-line sigma model, which enables the
non-perturbative frequency dependence of $\Pi_{ij}(\w)$ to be calculated
using a saddle-point expansion about a geometrically motivated classical 
solution and ~(ii) the Penrose plane-wave limit, which encodes the relevant 
tidal effects of the spacetime curvature in the neighbourhood of the original 
null geodesic traced by the photon in the classical theory.

\section{The World-Line Sigma Model, Penrose Limit and Null Congruences}

\begin{figure}[ht] 
\centerline{\includegraphics[width=2.5in]{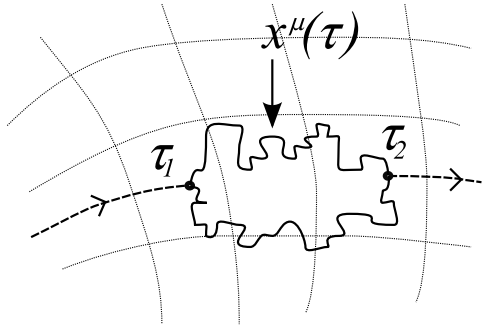}}
\caption{\footnotesize The $e^+e^-$ 
loop $x^\mu(\tau)$ with insertions of photon vertex
operators at $\tau_1$ and $\tau_2$.}\label{pic7}
\end{figure}

In the world-line formalism for scalar QED, the one-loop vacuum polarization 
tensor is given by\footnote{For relevant references
on the world-line technique \cite{Feynman:1950ir,Schwinger:1951xk} (for a
review, see \cite{Schubert:2001he}) in curved spacetime, 
see refs.\cite{Bastianelli:2002fv,Bastianelli:2003bg,Bastianelli:2005rc,
Kleinert:2002zn,Kleinert:2002zm}. We consider scalar QED for simplicity.
The extension to spinor QED involves the addition of a further Grassmann 
variable in the action \eqref{bb}, but introduces no new conceptual issues.}
\EQ{
\Pi^\text{1-loop}_{ij}=
\frac\alpha{4\pi}
\int_0^\infty\frac{dT}{T^3}\int_0^T d\tau_1\,d\tau_2\,{\cal Z}\,
\big\langle V^*_{\omega,\varepsilon_i}[x(\tau_1)]
V_{\omega,\varepsilon_j}[x(\tau_2)]\big\rangle\ .
\label{ba}
}
Here, $V_{\omega,\varepsilon_i}[x(\tau)]$ are vertex operators for the photon
and the expectation value is calculated in the 1-dim world-line sigma model 
involving periodic fields $x^\mu(\tau)=x^\mu(\tau+T)$ with an action 
\EQ{
S=\int_0^T d\tau\,\Big(\frac14g_{\mu\nu}(x)\dot x^\mu\dot x^\nu-m^2\Big)\ ,
\label{bb}
}
where $g_{\m\n}$ is the metric of the background
spacetime.\footnote{Since 
there is in general no translational invariance in curved spacetime,
the effective action depends on a point in spacetime and we choose our 
coordinates such that this is $x^\mu=0$. In the sigma model we deal with 
the corresponding zero mode in the ``string inspired'' way by imposing 
\cite{Bastianelli:2002fv,Bastianelli:2003bg,Bastianelli:2005rc,Kleinert:2002zn,Kleinert:2002zm,Schubert:2001he}  
$$
\int_0^T d\tau\,x^\mu(\tau)=0\ .
$$
This yields a translationally invariant formalism on the world-line that
allows us to fix $\tau_1=0$. We will then take $\tau_2=\xi T$, $0\leq\xi\leq 1$, 
and replace the two integrals over $\tau_1$ and $\tau_2$ by a single integral
over the variable $\xi$.} 
The expectation value is represented to ${\cal O}(\a)$ by an $e^+e^-$ loop 
with insertions of the photon vertex operators at $\t_1$ and $\t_2$, as 
illustrated in fig.~(\ref{pic7}). The factor ${\cal Z}$ is the partition 
function of the world-line sigma model relative to flat space. 

The form of the vertex operators is determined at this order by the classical 
equations of motion for the gauge field $A_\m(x)$ in the geometric optics,
or WKB, limit.\footnote{There are 3 dimensionful quantities in this problem:
the electron mass $m$, the frequency $\w$ of the photon, and the curvature scale 
$R$ (of mass dimension 2). We work in the WKB limit $\omega\gg \sqrt R$ in a 
weakly-curved background $R\ll m^2$. This leaves the dimensionless ratio 
$\omega^2 R/m^4$ to define the high and low frequency regimes.}
That is, 
\EQ{
V_{\omega,\varepsilon_i}[x] = \dot x^\m A_\m(x)
\label{bc}
}
where
\EQ{
A_\m(x) = {\cal A}(x) \hat\varepsilon_{i\m}(x)
e^{i\omega\Theta(x)}+\cdots\ ,
\label{bd}
} 
where ${\cal A}$ is the scalar amplitude, $\hat\varepsilon_i$ are the 
polarizations and the phase $\Theta$ satisfies the 
{\it eikonal equation\/}
\EQ{
g^{\mu\nu}\partial_\mu\Theta\partial_\nu\Theta=0\ .
\label{be}
}
The solution determines a {\it congruence\/} of null geodesics where
$\ell^\mu=\partial^\mu\Theta$ 
is the tangent vector to the null geodesic in the congruence
passing through the point $x$. In the particle interpretation,
$k^\mu=\omega\ell^\mu$ is momentum of a photon travelling along the geodesic.
The polarization vectors $\hat\varepsilon_i^\m$ are orthogonal to $\ell^\m$ 
and are parallel transported: $\ell\cdot D\,\hat\varepsilon_i^\mu=0$,
while the scalar amplitude satisfies 
\EQ{
\ell\cdot D\,\log{\cal A} = -{1\over2}D^\m \ell_\m \equiv -\hat\theta
\label{bg}
}
where the {\it expansion\/} 
$\hat\theta$ is one of the optical scalars appearing in
the Raychoudhuri equations.

In order to evaluate the world-line path integral over 
$x^\m(\t)$, we will need to consider fluctuations about the classical
geodesic $\c$. The first step is to set up Fermi normal coordinates
$(u,v,y^i)$ \cite{Blau:2006ar}
which are adapted to the null geodesic in the sense that
$\c$ is the curve $(u,0,0,0)$ where $u$ is the affine parameter, $v$ is
another null coordinate and $y^i$ parameterize the transverse subspace.
Now, as explained in Section 3, the relevant curvature degrees of freedom 
needed to describe these fluctuations at leading order in an expansion
in $R/m^2$ are captured in the Penrose limit of the spacetime around $\c$
\cite{pen,Blau2}.
This follows from an overall Weyl re-scaling $ds^2 \to \l^2 ds^2$ of the 
metric obtained by an asymmetric re-scaling of the coordinates,
$(u,v,y^i) \to (u,\l^2 v,\l y^i)$, chosen so that the affine parameter 
$u$ is invariant. This leads, for an arbitrary background spacetime,
to the plane wave metric
\EQ{
ds^2=2du\,dv+h_{ij}(u)y^i\,y^j\,du^2-dy^{i2}\ ,
\label{bh}
}
where $h_{ij}(u)$ is related to the curvature of the original metric 
$g_{\m\n}$ by $h_{ij} = -R_{uiuj}$. This is the Penrose limit of the
background spacetime in the neighbourhood of $\c$ in Brinkmann 
coordinates.

\begin{figure}[ht] 
\centerline{(a)\includegraphics[width=2.5in]{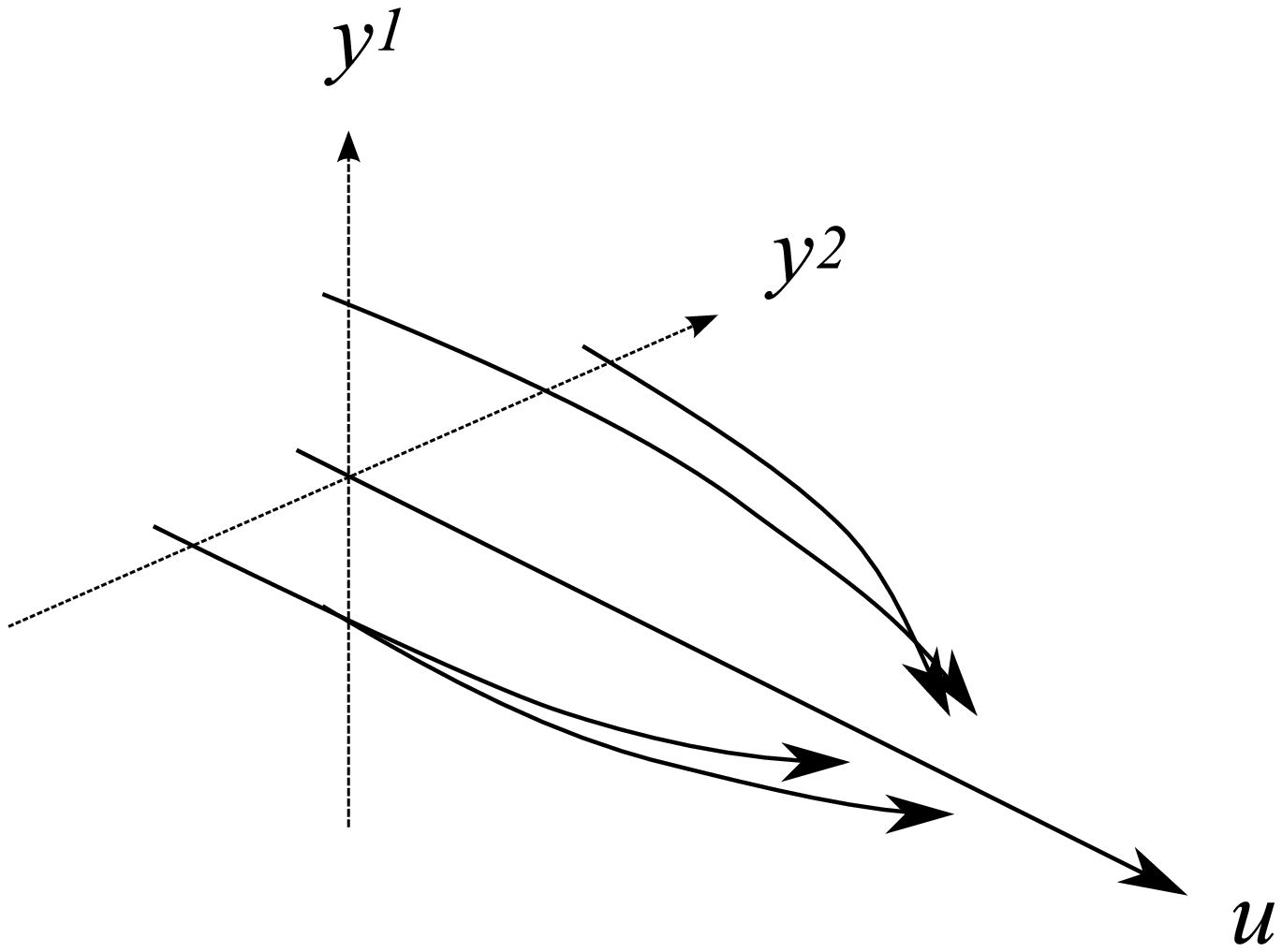}\hspace{0.2cm}
(b)\includegraphics[width=2.5in]{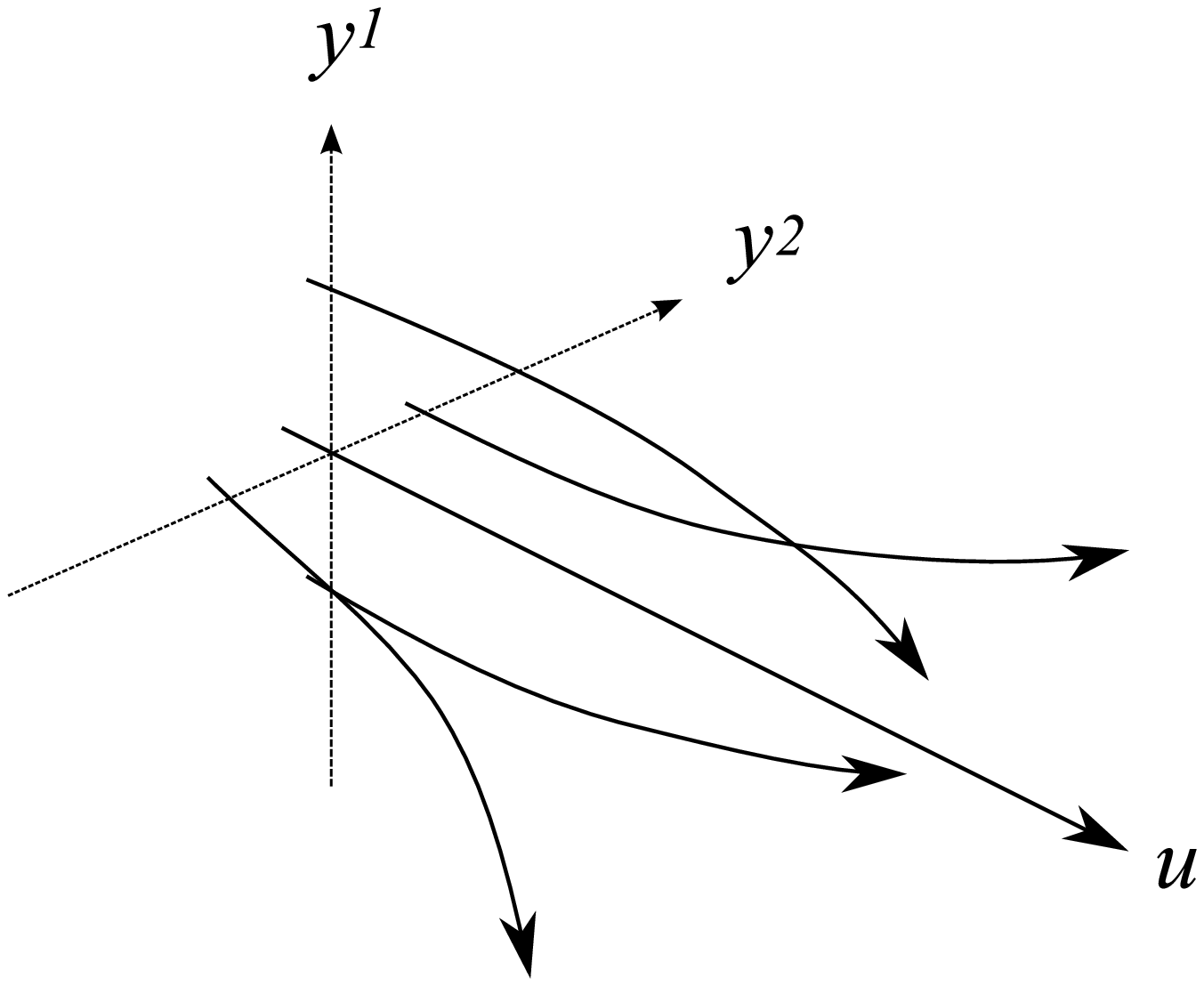}}
\caption{\footnotesize The behaviour of the congruence of null
geodesics for the (a) Type I and (b) Type II plane wave metrics.} 
\label{pic1}
\end{figure}

It is now convenient to divide these spacetimes into two classes, depending
on the behaviour of the null congruence. Introducing a second optical
scalar, the shear $\hat\s =\sqrt{\tfrac12D_{(\m}\ell_{\n)} D^\m \ell^\n - 
\hat\theta^2}$, we can write the Raychoudhuri equations in
the form:
\SP{
\partial_u({\hat\theta} + {\hat\s}) &= - (\hat\theta + \hat\s)^2
-\Phi_{00} -|\Psi_0|\ ,\\
\partial_u({\hat\theta} - {\hat\s}) &=- (\hat\theta - \hat\s)^2
- \Phi_{00} + |\Psi_0|\ .
\label{bi}
}
Here, we have introduced the Newman-Penrose notation for the components 
of the Ricci and Weyl tensors: 
$\Phi_{00} = {1\over2} R_{uu}$ and
$|\Psi_0| = {1\over2}(C_{u1u1}-C_{u2u2})$.
The effect of expansion and shear is visualized by the effect on a 
circular cross-section of the null congruence as the affine parameter $u$
is varied: the expansion $\hat\theta$ gives a uniform expansion whereas the
shear $\hat\s$ produces a squashing with expansion along one transverse axis
and compression along the other. The combinations $\hat\theta \pm \hat\s$ 
therefore describe the focusing or defocusing of the null rays in the two
orthogonal transverse axes. Provided the signs of $\Phi_{00} \pm |\Psi_0|$
remain fixed (as in the symmetric plane wave example 
considered in detail later)
we can therefore divide the plane wave metrics into two classes 
(see fig.~(\ref{pic1})).
A Type I spacetime, where $\Phi_{00} \pm|\Psi_0|$ are both positive, 
has focusing in both directions, whereas Type II,
where $\Phi_{00} \pm |\Psi_0|$ have opposite signs, has one focusing and
one defocusing direction. Note, however, that there is no spacetime with
both directions simultaneously defocusing, since the null-energy condition 
requires $\Phi_{00} \ge 0$.

It is clear that provided the geodesics are complete, those in a focusing 
direction will eventually cross. In fact, the existence of 
conjugate points, as described earlier,
is generic in spacetimes satisfying the null 
energy condition \cite{HawkingEllis,Wald}. 
The existence of conjugate points plays a crucial r\^ole in the world-line 
path integral formalism since, as explained below, they imply the existence
of zero modes in the partition function which ultimately are responsible
for the Kramers-Kronig violating singularities in the vacuum polarization 
tensor. 

First, we need the explicit solutions for the geodesic equations in the 
plane wave metric \eqref{bh}. With $u$ itself as the affine parameter,
these are:
\EQ{
\partial_u^2v +2 h_{ij} y^i \partial_u y^j + 
{1\over2}\partial_uh_{ij} y^i y^j = 0 \ ,\qquad
\partial_u^2y^i + h_{ij} y^j = 0\ ,
\label{bj}
}
the latter being the Jacobi equation for the geodesic deviation vector $y^i$.
The solution for $v$ determines the eikonal phase \cite{us}
\EQ{
\Theta(x) = v - {1\over2} \Omega_{ij}y^i y^j\ ,
\label{bk}
}
where $\Omega_{ij} = \partial_uE_{ia}E^a{}_j$ is most simply expressed in terms
of a zweibein\footnote{The zweibein $E^a{}_i(u)$ and its inverse
  $E^i{}_a(u)$ relate the
Brinkmann coordinates $y^i$ used here to Rosen coordinates $Y^a$, which are
particularly well-suited to describing the null congruence but not so
simple for evaluating the world-line path integral \cite{us}.
A particular geodesic in the congruence has $y^i = E^i{}_a Y^a$ and
$v = \Theta + {1\over2} \Omega_{ij}y^i y^j$ for constant $Y^a,
\Theta$. In our conventions $i$ and $j$ in this 2d Euclidean subspace
are raised and lowered with $\delta_{ij}$.}
$E^a{}_i(u)$
in terms of which the curvature is $h_{ij} = -\partial_u^2E_{ia}E^a{}_j$.
The polarization 1-forms and scalar amplitude are
\EQ{
\hat\varepsilon_i = dy^i -\Omega_{ij} y^j du \ ,\qquad
{\cal A} = \sqrt{1\over {\rm det} E^i{}_a}\ . 
\label{bm}
}
These give all the ingredients necessary for the vertex operators, so we
determine
\EQ{
V_{\omega,\varepsilon_i}[x^\mu(\tau)]=
\big(\dot y^i -\Omega_{ij} y^j \dot u \big) ~
\sqrt{1\over {\rm det} E^i{}_a}~
\exp i\omega\Big[v - {1\over2} \Omega_{ij} y^i y^j\Big]\ .
\label{bn}
}

\section{World-line Calculation of the Vacuum Polarization}

We have now assembled all the elements needed to calculate the world-line
path integral \eqref{ba} for the vacuum polarization. 
The fundamental idea is to evaluate this by considering the Gaussian
fluctuations about a saddle point given by the classical solution of the
equations of motion for the action
\EQ{
S=-T+\frac{m^2}{4T}\int_0^1d\tau\,g_{\mu\nu}(x)\dot x^\mu\dot x^\nu-
\omega\Theta[x(\xi)]+\omega\Theta[x(0)]\ ,
\label{ca}
}
{\it including} the phase $\Theta(x)$ of the vertex operators which act as
sources. In \eqref{ca}, we have re-scaled $T \to T/m^2$, which makes it 
clear that $1/m^2$ plays the r\^ole of a conventional coupling constant.
In fact, the effective dimensionless coupling is actually the dimensionless
ratio $R/m^2$. (We have also re-scaled $\t \to T\t$ so that the $\t$ integral 
now runs from 0 to 1.)

The sources act as impulses which insert world-line momentum at the 
special points $x(0)$ and $x(\xi)$. In between these points, the classical
solution is simply a null geodesic path and it is straightforward to see
that this is given by $v = y^i = 0$ while $u$ satisfies
\EQ{
\ddot u=-\frac{2\omega T}{m^2}\delta(\tau-\xi)+
\frac{2\omega T}{m^2}\delta(\tau)\ .
\label{cb}
}
The solution is \cite{us}
\EQ{
u = \tilde u(\tau)=-u_0+\begin{cases}2\omega T(1-\xi)\tau/m^2 & 0\leq 
\tau\leq
\xi\ ,\\ 2\omega T\xi(1-\tau)/m^2 & \xi\leq\tau\leq 1\ ,\end{cases}
\label{cc}
}
where $u_0=\omega T\xi(1-\xi)/m^2$. This describes an $e^+e^-$ loop
squashed down to lie along the original photon null geodesic
between $u = \pm u_0$ as illustrated in fig.~(\ref{pic17}).

An intriguing aspect of this is that the affine parameter length of the
loop actually {\it increases} as the frequency $\w$ is increased. 
Technically, this is simply because for higher frequencies the sources
impart larger impulses. However, it means that in the high frequency limit,
the photon vacuum polarization probes the entire length of the null geodesic
and becomes sensitive to global aspects of the null congruence.
This is rather counter-intuitive, since we would naively expect high
frequencies to probe only local regions of spacetime, and appears to be yet 
another example of the sort of UV-IR mixing phenomenon seen in other
contexts involving quantum gravity or string theory.

Another crucial point is that while \eqref{cc} is the only general solution,
for specific values of $T$ (or equivalently $\w$) there are further
solutions. These are due to the existence of conjugate points on the null
geodesic. For values of $T$ for which $v(\pm u_0) = y^i(\pm u_0) = 0$, there 
exists more than one null geodesic path between the points $\t = 0,\xi$
where the impulses are applied. This gives rise to a continuous set of
classical solutions, which results in zero modes in the path integral
for these specific values of $T$: see fig.~(\ref{pic17}). 
In turn, these produce the singularities 
in the partition function which are responsible for the violation
of  the Kramers-Kronig relation. 
Notice, that it is not necessary that these more
general geodesics lift from the Penrose limit to the full metric.

The perturbative expansion in $R/m^2$ about this solution can be 
made manifest by an appropriate re-scaling of the ``fields'' $x^\m(\t)$. 
However, this re-scaling must be done in such a way as to leave the 
classical solution $\tilde u(\t)$ invariant. But this is precisely
(see ref.\cite{us} for details) the Penrose re-scaling described above,
where $\l$ is identified with the effective coupling $R/m^2$. This is why 
the physics of vacuum polarization is captured perturbatively in 
${\cal O}(R/m^2)$ by the Penrose expansion in ${\cal O}(\l)$ 
\cite{Blau:2006ar,us} of the background spacetime around the photon's
null geodesic $\c$.

Expanding around the classical solution, in the plane wave metric \eqref{bh},
we find the Gaussian fluctuations in the transverse directions (the $u$,
$v$ fluctuations are identical to flat space) are governed by the action
\EQ{
S^{(2)}=\int_0^1d\tau\,\Big[-\frac1{4}\big(
\dot y^{i2}+\dot{\tilde u}^2 h_{ij}(\tilde u)y^i y^j\big)
+\frac{\omega T}{2m^2}
\Omega_{ij}(\tilde u)y^i y^j \big(\d(\t - \xi) - \d(\t)\big)
\Big]\ .
\label{cd}
}

At this point, in order to illustrate the general features of our analysis
with a simple example, we restrict the plane wave background to the special
case where the curvature tensor is covariantly constant. This defines
a {\it locally symmetric} spacetime, and the corresponding ``symmetric
plane wave'' has metric \eqref{bh} with
\EQ{
h_{ij} y^i y^j = \s_1^2 (y^1)^2 + \s_2^2 (y^2)^2\ .
\label{ce}
}
The curvatures are $R_{uu} = \s_1^2 + \s_2^2$ and $C_{u1u1} = -C_{u2u2} =
{1\over2}(\s_2^2 - \s_1^2)$. The signs of the coefficients determine which 
class the spacetime belongs to. For Type I (focus/focus), $\s_1$ and $\s_2$
are both real, while Type II (focus/defocus) has $\s_1$ real and $\s_2$
imaginary. The case $\s_1,\s_2$ both imaginary is forbidden by the null
energy condition, which requires $R_{uu} > 0$.
With this background, the geodesic equations determine the 
zweibein $E^i{}_a{} =\delta_{ai}\cos(\sigma_iu+a_i)$ and $\Omega_{ij}
=-\delta_{ij}\s_i\tan(\s_iu+a_i)$ (the $a_i$ are arbitrary integration
constants).

The core of the calculation reported in ref.\cite{us} is the evaluation
of the partition function ${\cal Z}$ and the two-point function
$\langle \hat\varepsilon_i \cdot \dot x ~ 
\hat\varepsilon_j \cdot \dot x\rangle$ for the world-line path integral
\eqref{ba} in the symmetric plane wave background. A lengthy calculation
described in detail in \cite{us} eventually gives the following result
for the vacuum polarization:
\EQ{
\Pi^\text{1-loop}_{ij}
=-\delta_{ij}\frac{\alpha m^2}{2\pi}\int_0^{\infty+i\epsilon} \frac{dT}{T^2}
e^{-T}\int_0^1d\xi\,\left\{
1-\frac{\beta_i}{\sinh\beta_i\cosh\beta_i}
\prod_{l=1}^2
\sqrt{\frac{\beta_l^3}{\sinh^3\beta_l\cosh\beta_l}}\right\}\ .
\label{cf}
}
where $\beta_i=\omega T\xi(1-\xi)\sigma_i/m^2$. In deriving \eqref{cf}, 
we have performed a Wick rotation $T\to-iT$ to leave a convergent
integral. The $i\epsilon$ prescription deals with the singularities on
the real axis which arise in the Type II case.

\begin{figure}[ht] 
\centerline{\includegraphics[width=2.5in]{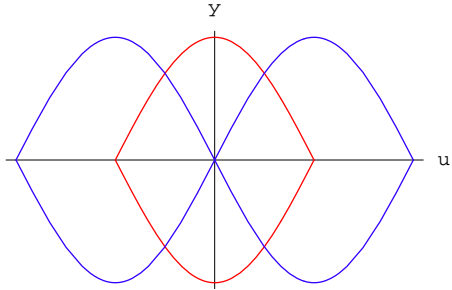}}
\caption{\footnotesize The $n=1$ (red) and $n=2$ (blue) zero modes for
  $\xi=\tfrac12$. The points $u=\pm u_0$ are conjugate points for 
$\gamma$.}\label{plot6}
\end{figure}
This expansion demonstrates clearly the unconventional analyticity
properties of the vacuum polarization in curved spacetime.
When $\s_i$ is real, there are branch point singularities on the imaginary
$T$ axis at the specific values 
\EQ{
T=\frac{i\pi m^2n}{2\xi(1-\xi)\sigma_i\omega}\ ,\qquad n=1,2,\ldots\ .
\label{cg}
}
These singularities arise from zeros of the
fluctuation determinant ${\cal Z}$ and have a natural interpretation
in terms of zero modes, {\it i.e.\/}~non-trivial solutions of the
fluctuation equations that follow by varying \eqref{cd} with respect
to $y^i$. For the special case $\xi=\tfrac12$, these zero modes are 
particularly simple to write down: $u=\tilde u(\tau)$ as in \eqref{cc} 
while
\EQ{
y^i(\tau)=\sin(2n\pi\tau)\ .
\label{ch}
}
The $n=1$ and $n=2$ zero modes are illustrated in fig.~(\ref{plot6}). 
(For generic $\xi$, the zero modes are more complicated: see ref.\cite{us}.)
This confirms that the zero modes are associated to geodesics that intersect
$\gamma$ at both $u=\pm u_0$, {\it i.e.\/}~the conjugate points on the 
geodesic $\gamma$. In terms of the vacuum polarization 
$\Pi_{ij}^{\text{1-loop}}(\w)$ or the refractive index $n_i(\w)$, these 
singularities appear on the imaginary axis in the upper-half plane in $\w$.
They have profound consequences for the Kramers-Kronig dispersion relation. 

In the Type II case, where one of the $\s_i$ is imaginary, there are also
singularities on the real $T$ axis. These singularities can also be 
understood in terms of zero modes, but now with imaginary affine parameter 
$u\to iu$, and can be thought of as world-line instantons. These give rise to 
an imaginary part for $\Pi_{ij}^{\text{1-loop}}(\w)$ and the refractive index
$n(\w)$, which corresponds to the tunnelling process $\c \to e^+e^-$
in the background gravitational field.

\section{Refractive Index in Curved Spacetime}

The refractive index follows immediately from the result \eqref{cf} for the
vacuum polarization. We present the results for the Type I and Type II 
symmetric plane wave backgrounds separately.

\begin{figure}[ht] 
\centerline{(a)\includegraphics[width=2.5in]{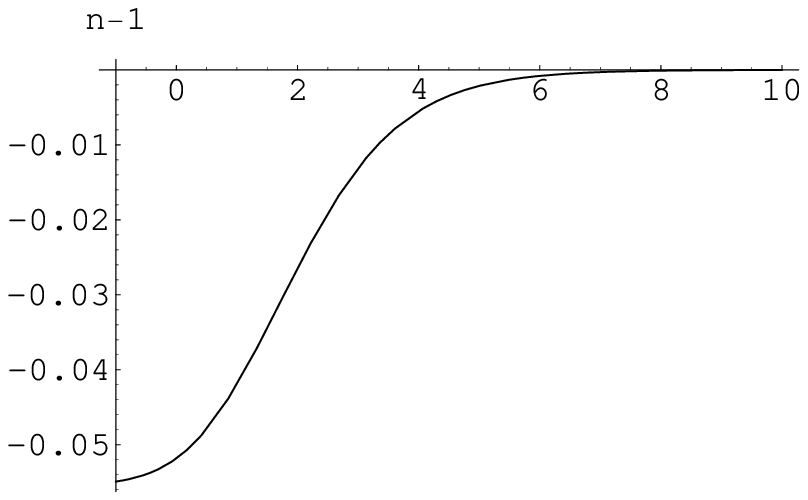}\hspace{0.2cm}
(b)\includegraphics[width=2.5in]{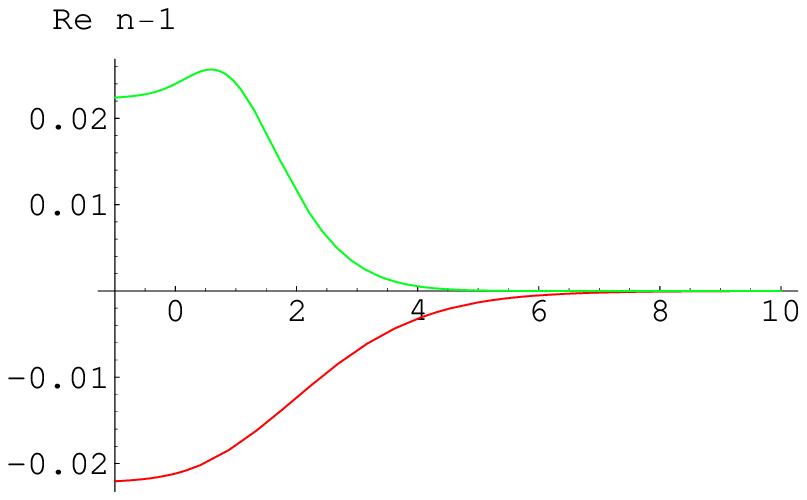}}
\caption{\footnotesize The behaviour of
$\text{Re}\,n_i(\omega)-1$ in
  units of $\alpha R/(2\pi m^2)$, as a
  function of $\tfrac12\log\omega^2 R/m^4$ for (a) Type I conformally 
  flat case ($n_1=n_2$) and (b) 
Type II Ricci flat case  ($i=1$ red, $i=2$ green). Note that in both cases
$n_i(\w)$ approaches 1 from below as $\w \to \infty$.} 
\label{plot1}
\end{figure}

\noindent{\it Type I:}~~~This includes the special case of conformally flat spacetimes, where $\s_1 = \s_2 \equiv\sqrt R$.  Evaluating \eqref{cf} numerically gives the result shown in fig.~(\ref{plot1}).
Here, both polarizations are superluminal at low frequencies, with the
refractive index rising monotonically to the high-frequency limit $n(\w)\to1$.
The wavefront velocity, $v_{\rm wf} = v_{\rm ph}(\infty)$ is therefore $c$,
in accordance with our expectations from causality.  The integrand in 
\eqref{cf} is regular on the real axis and so $\text{Im}~n(\w)$ is
vanishing and there is no pair creation.

We can find explicit analytic expressions for $n(\w)$ in these limits \cite{us}.
For low frequencies, 
\EQ{
n_i(\omega)=1-\frac{\alpha R}{2\pi m^2}\Big[\frac1{18}
-\frac{71}{14175}\frac{\omega^2R}{m^4}+
\frac{428}{189189}\Big(\frac{\omega^2R}{m^4}\Big)^2-\cdots\Big]\ .
\label{da}
}
This series is divergent but alternating and this is correlated
with the fact that it is Borel summable, with the sum being defined by the 
convergent integral in \eqref{cf} which has no singularities on the real axis. 
In the high-frequency limit, we find that the refractive index approaches 1
from below, with a $1/\w$ dependence:
\EQ{
n_i(\omega)=
1-\frac{\alpha C_i}{12\pi\omega}+{\cal O}\Big(\frac{\log\omega}{\omega^2}\Big)
\,
\label{db}
}
where $C_i=\big(\frac13+\frac{7\pi^2}{36}\big)\sqrt R$
for both $i=1,2$.

\begin{figure}[ht] 
\centerline{(a)\includegraphics[width=2.5in]{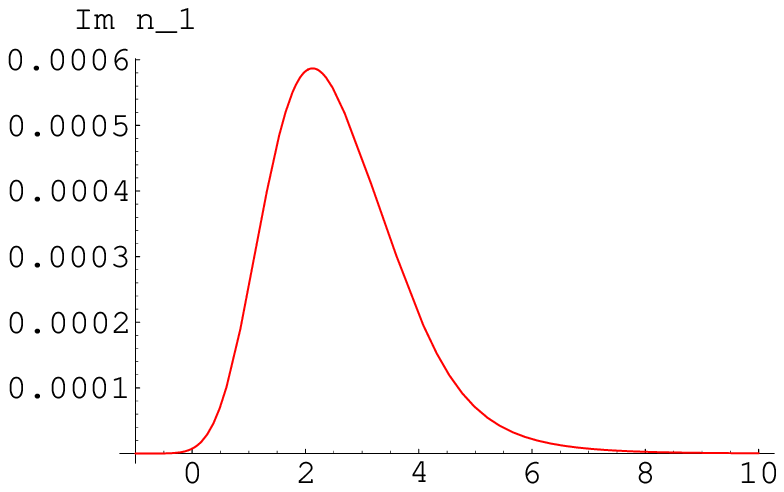}
\hspace{0.2cm}(b)\includegraphics[width=2.5in]{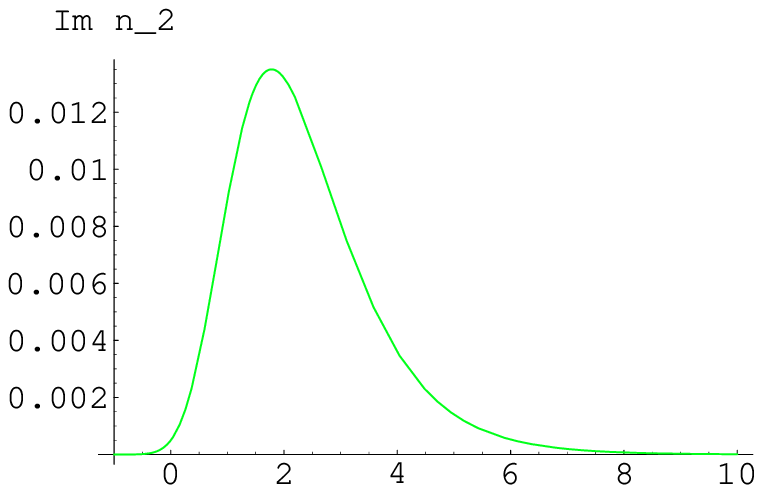}}
\caption{\footnotesize The behaviour of
(a) $\text{Im}\,n_1(\omega)$ and 
(b) $\text{Im}\,n_2(\omega)$ in
  units of $\alpha R/(2\pi m^2)$, as a
  function of $\tfrac12\log\omega^2 R/m^4$ for the Type II Ricci flat 
case.}\label{plot2}
\end{figure}

\noindent{\it Type II:}~~~In this case, the integrand in \eqref{cf} has branch
point singularities on both the real and imaginary axes. 
The refractive indices therefore have both a real and imaginary part. 
These are shown figs.~(\ref{plot1}) and (\ref{plot2}) for the special
case of a Ricci flat background with $\sigma_1=i\sigma_2=\sqrt R$. 
Notice that the
subluminal polarization state $n_2(\omega)$ has the same general form
as the refractive index for a conventional absorptive optical medium 
\cite{Shore:2007um,us}.

In this case the propagation displays {\it gravitational bi-refringence\/}
\cite{Drummond:1979pp,Shore:1995fz} since the two polarizations have different phase velocities.
In general, the low-frequency limit of the refractive index is
\EQ{
n_i(\omega)=1-\frac{\alpha}{360\pi}{1\over m^2}\big(
10\Phi_{00}\mp 4|\Psi_0|\big)+{\cal O}(\omega^2)\ .
\label{dd}
}
for $i = 1,2$.
For Type II, at low frequencies, we find 
has
\EQ{
n_{1,2}(\omega)=1\mp\frac{\alpha R}{2\pi m^2}\Big[\frac1{45}
\mp\frac{37}{28350}\frac{\omega^2 R}{m^4}+
\frac{34}{85995}\Big(\frac{\omega^2R}{m^4}\Big)^2 \mp \cdots\Big]\ .
\label{de}
}
for the superluminal and subluminal polarizations respectively.
In both cases the Borel
transforms have branch point singularities on the real axis and this is
indicative of an imaginary part which vanishes to all orders in the 
$\omega^2R/m^4$ expansion. In fact, for low frequencies, 
$\text{Im}\,n(\omega)$ is dominated by the closest singularity to the
origin, leading to the universal behaviour 
\EQ{
\text{Im}\,n_i(\omega)\thicksim
\exp-\frac{2\pi m^2}{\omega|\sigma_2|}\ .
\label{dg}
}
The high-frequency limit is again given by \eqref{db}, where this time
the $C_i$ are complex but still with $\text{Re}~C_i >0$, so that all 
polarizations for both Type I and Type II have phase velocities that approach
$c$ at high frequencies from the superluminal side.\footnote{
All this shows clearly that the most important frequency dependence of 
$n(\w)$ is non-perturbative in the parameter $\w^2 R/m^4$.
It was therefore not captured by previous effective action approaches
\cite{Shore:2002gn}, which evaluated all orders in a derivative expansion
but were restricted to ${\cal O}(R/m^2)$. The necessity for a non-perturbative
technique to determine $n(\w)$ was already noted in refs.\cite{Shore:2002gn,
Shore:2003jx}.}

\section{Micro-Causality and the Kramers-Kronig Relation}

\begin{figure}[ht]
\centerline{\includegraphics[width=2.5in]{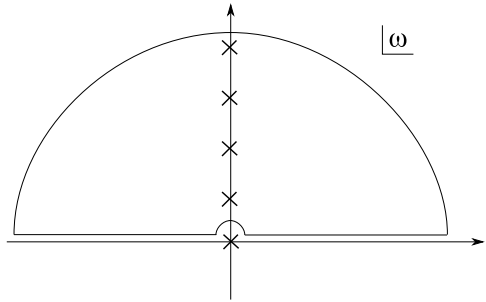}}
\caption{\footnotesize The integration contour for $\oint
  d\omega\,n(\omega)/\omega$ used in the proof of the generalized KK relation 
  for the conformally flat symmetric wave with poles on the imaginary axis.}
\label{pic15}
\end{figure}

These results on the analyticity structure of $\Pi^{\text{1-loop}}(\w)$
and $n(\w)$ explain why the Kramers-Kronig dispersion relation fails to hold
for QED in curved spacetime. The derivation of \eqref{aa} relies on the fact
that $n(\w)$ is analytic in the upper-half plane. As we have seen, however, 
this is not true in curved spacetime because $n(\w)$ has singularities
on the imaginary axis (see fig.~(\ref{pic15})) and these must be included in
the contour integral used to derive \eqref{aa}.
So, for example, in the conformally flat Type I case, the
singularities are poles and we have
\SP{
&\int_\text{semi-circle}\frac{d\omega}\omega\,n(\omega)
-\pi i n(0)+{\cal P}\int_{-\infty}^\infty \frac{d\omega}\omega\,
n(\omega)\\
&=\pi i\big(n(\infty)-n(0)\big)+
{\cal P}\int_{-\infty}^\infty \frac{d\omega}\omega\,
n(\omega) ~=~ \text{pole contribution}\ .
\label{ea}
}
In this case $\text{Im}\,n(\omega)=0$ and 
the principal value integral vanishes, 
and \eqref{ea} becomes
\EQ{
\text{Re}\,n(0)-\text{Re}\,n(\infty)=\frac{\alpha R}{\pi m^2}
\int_0^\infty dT\,e^{-T}\,\int_0^1d\xi\,\big(\xi(1-\xi)\big)^2\,
\sum_{n=1}^\infty \text{Res}\,f(i\pi n/2)\ ,
\label{eb}
}
where $f(x)=\big(1-\tfrac{x^4}{\sinh^4x\cosh^2x}\big)\tfrac1{x^3}$. Performing the residue
sum we find
\EQ{
\text{Re}\,n(0)-\text{Re}\,n(\infty)=-\frac{\alpha R}{36\pi m^2}\ ,
\label{ec}
}
which is in perfect agreement with \eqref{da} and \eqref{db}.

The fact that $n(\w)$ is not analytic in the upper-half plane
calls into question the validity of 
micro-causality. Recall that in relativistic QFT, micro-causality,
{\it i.e.\/}~the vanishing of commutators of field operators for spacelike
separations, implies that the retarded propagator $\Delta_{\rm ret}$ is only
non-vanishing in, or, as in the present case of a massless quantum, on, 
the forward null cone. At tree level, this remains
true for QED in curved spacetime. However, at one loop, the vacuum polarization
$\Pi_{ij}^{\text{1-loop}}$ contributes to the full propagator:
$\Delta=\Delta^\text{tree}-\Delta^\text{tree}
\Pi^\text{1-loop}\Delta^\text{tree}+\cdots$, 
and we must check whether $\Pi^\text{1-loop}$ 
itself only has support in/on the forward cone.

{}From our calculation of the on-shell momentum-space vacuum polarization 
tensor $\Pi^{\text{1-loop}}(\w)$, we can attempt to determine the 
dependence of the 
real-space $\Pi^{\text{1-loop}}$ on the null coordinate $v$ by taking 
a Fourier transform:
\EQ{
\Pi^\text{1-loop}(v)=
\int_{-\infty}^{\infty} d\omega\,e^{-i\omega v}\Pi^\text{1-loop}(\omega)
\ .
\label{ed}
}
This is a retarded quantity if the integration contour is taken to 
avoid singularities by veering into the upper-half plane,
when $v<0$, and the lower half plane, when $v>0$. For QFT in flat spacetime,
$\Pi^\text{1-loop}(\omega)$ is analytic in the upper-half plane and so
when $v<0$ one computes the $\omega$ integral by completing the contour with a
semi-circle at infinity in the upper-half plane. Since there are no
singularities in the upper-half plane, the integral
vanishes and consequently $\Pi^\text{1-loop}_\text{ret}(v)=0$ for
$v<0$. This is consistent with the fact that
the region $v<0$ lies outside the forward light
cone. Hence, in this case micro-causality is preserved as a
consequence of analyticity in the upper-half plane in frequency space.

\begin{figure}[ht] 
\centerline{\includegraphics[width=2.2in]{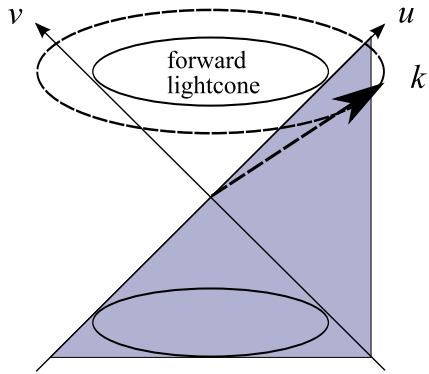}}
\caption{\footnotesize Including vacuum polarization effects, the
  photon momentum $k$ may lie outside the forward light cone ($u>0$,
  $v>0$) of its original null geodesic $v=0$. The potential 
  violation of
  micro-causality arises if the retarded propagator is
  non-vanishing even for $v < 0$ (the shaded area), which
  lies outside the forward light cone.}\label{pic14}
\end{figure}

In curved spacetime, on the
contrary, $\Pi^\text{1-loop}(\omega)$ is not
analytic in the upper-half plane and consequently 
$\Pi^\text{1-loop}_\text{ret}(v)$ may 
receive contributions from the region $v<0$ which lies 
outside the forward light cone. See fig.~(\ref{pic14}).
Indeed, for the conformally flat case where 
$\Pi^\text{1-loop}_\text{ret}(\w)$ has poles on the imaginary axis,
we can estimate the behaviour of $\Pi^\text{1-loop}_\text{ret}(v)$
as
\EQ{
\Pi^\text{1-loop}_\text{ret}(v)
\thicksim \exp-\sqrt{\frac{8\pi m^2|v|}{R^{1/2}}}\ .
\label{ef}}
which appears to show a violation of micro-causality with an
exponential dependence on a characteristic time/length scale $\sqrt R/m^2$. 

However, in order to be certain that this is a genuine property of the
full real-space propagator, it is necessary to calculate the vacuum
polarization {\it off-shell} \cite{Dubovsky:2007ac}. In our ${\cal O}(\a)$
on-shell calculation of $\Pi_{ij}$, the frequency $\w$ is identified with 
the light-cone component $p_+$, while the component $p_-$ is taken to zero.
It remains possible that when $p_-$ is small, but non-vanishing,
the non-analyticities are shifted into the causally safe region
$\text{Im}\,p_+/\text{Im}\,p_-<0$.\footnote{{\it Note added}: We have now 
completed a full off-shell calculation of the vacuum polarization tensor
and find that precisely this behaviour occurs. Full details will be presented
elsewhere \cite{usagain}.}

It will be especially interesting 
to explore these issues of superluminal propagation and causality 
in spacetimes such as Schwarzschild with both
horizons and singularities. Although both $\Phi_{00}$ and $\Psi_0$ vanish
for principal null geodesics at an event horizon, higher-order terms in the 
Penrose expansion do play a r\^ole and the non-vanishing of commutators in the
neighbourhood of a horizon could have profound consequences. The UV-IR effect
whereby high frequencies probe the global properties of the photon's null 
geodesic could be especially important in spacetimes with singularities.
We have implicitly assumed here that the null geodesics in the congruence
are complete. However, this is no longer true in the presence of singularities,
raising the intriguing possibility that their existence could affect the 
high frequency behaviour of photon propagation in a global rather than purely 
local way.  

\vskip0.5cm
We would like to thank Asad Naqvi for many useful conversations and 
Sergei Dubovsky, Alberto Nicolis, Enrico Trincherini and 
Giovanni Villadoro for pointing out the necessity of working
off-shell in order to completely settle the question of micro-causality.
TJH would also like to thank Massimo Porrati for a helpful discussions
and Fiorenzo Bastianelli for explaining some details 
of his work on the world-line formalism. This work was supported in part
by PPARC grant PP/D507407/1.

\end{document}